\documentclass[%
 reprint,
superscriptaddress,
 amsmath,amssymb,
 aps,
pra,
floatfix,
]{revtex4-2}

\usepackage{graphicx}
\usepackage[dvipsnames]{xcolor}
\usepackage{dcolumn}
\usepackage{bm}
\usepackage{tensor}
\usepackage{gensymb}
\usepackage{xr-hyper} 
\usepackage{hyperref}
\usepackage[version=4]{mhchem}
\usepackage{siunitx}
\usepackage[caption=false]{subfig}
\usepackage{derivative}
\usepackage{tensor}

\renewcommand{\selectlanguage}[1]{}

\DeclareSIUnit\gauss{G}
\DeclareSIUnit\mbar{mbar}

\DeclareMathVersion{nxbold} 
\SetSymbolFont{operators}{nxbold}{OT1}{cmr} {b}{n}
\SetSymbolFont{letters}  {nxbold}{OML}{cmm} {b}{it}
\SetSymbolFont{symbols}  {nxbold}{OMS}{cmsy}{b}{n}

\makeatletter
\newcolumntype{B}[3]{>{\boldmath\DC@{#1}{#2}{#3} }c<{\DC@end} }
\newcolumntype{Z}[3]{>{\mathversion{nxbold}\DC@{#1}{#2}{#3} }c<{\DC@end} } 
\newcommand*{\addFileDependency}[1]{
\typeout{(#1)}
\@addtofilelist{#1}
\IfFileExists{#1}{}{\typeout{No file #1.}}
}\makeatother

\begin{document}

\preprint{APS/123-QED}

\title{A Magneto-Optical Trap of Titanium Atoms}

\author{Scott Eustice}%
    \email{scott\_eustice@berkeley.edu}
    \affiliation{Department of Physics, University of California, Berkeley, CA 94720}
    \affiliation{Challenge Institute for Quantum Computation, University of California, Berkeley, CA 94720}
    \affiliation{Quantum Systems Accelerator, Lawrence Berkeley National Lab, Berkeley, CA 94720}
\author{Jackson Schrott}
    \affiliation{Department of Physics, University of California, Berkeley, CA 94720}
    \affiliation{Challenge Institute for Quantum Computation, University of California, Berkeley, CA 94720}
    \affiliation{Quantum Systems Accelerator, Lawrence Berkeley National Lab, Berkeley, CA 94720}
\author{Anke St\"oltzel}
    \affiliation{Department of Physics, University of California, Berkeley, CA 94720}
    \affiliation{Challenge Institute for Quantum Computation, University of California, Berkeley, CA 94720}
    \affiliation{Quantum Systems Accelerator, Lawrence Berkeley National Lab, Berkeley, CA 94720}
\author{Julian Wolf}
    \altaffiliation[Present address: ]{Vector Atomic Inc., Pleasanton, CA 94566}
    \affiliation{Department of Physics, University of California, Berkeley, CA 94720}
    \affiliation{Challenge Institute for Quantum Computation, University of California, Berkeley, CA 94720}
    \affiliation{Quantum Systems Accelerator, Lawrence Berkeley National Lab, Berkeley, CA 94720}
\author{Diego Novoa}
    \altaffiliation[Present address: ]{Vector Atomic Inc., Pleasanton, CA 94566}
    \affiliation{Department of Physics, University of California, Berkeley, CA 94720}
    \affiliation{Challenge Institute for Quantum Computation, University of California, Berkeley, CA 94720}
    \affiliation{Quantum Systems Accelerator, Lawrence Berkeley National Lab, Berkeley, CA 94720}
\author{Kayleigh Cassella}
    \altaffiliation[Present address: ]{Atom Computing Inc., Berkeley, CA 94710}
    \affiliation{Department of Physics, University of California, Berkeley, CA 94720}
    \affiliation{Challenge Institute for Quantum Computation, University of California, Berkeley, CA 94720}
    \affiliation{Quantum Systems Accelerator, Lawrence Berkeley National Lab, Berkeley, CA 94720}
\author{Dan M. Stamper-Kurn}
    \affiliation{Department of Physics, University of California, Berkeley, CA 94720}
    \affiliation{Challenge Institute for Quantum Computation, University of California, Berkeley, CA  94720}
    \affiliation{Quantum Systems Accelerator, Lawrence Berkeley National Lab, Berkeley, CA 94720}
    \affiliation{Materials Science Division, Lawrence Berkeley National Laboratory, Berkeley, CA 94720}

\date{\today}

\begin{abstract}
We realize laser cooling and trapping of titanium (Ti) atoms in a mangeto-optical trap (MOT). While Ti does not possess a transition suitable for laser cooling out of its 3d$^2$4s$^2$ a$^3$F ground term, there is such a transition, at an optical wavelength of $\lambda=\qty{498}{\nano\meter}$, from the long-lived  3d$^3$($^4$F)4s a$^5$F$_5$ metastable state to the 3d$^3$($^4$F)4p y$^5$G$^o_6$ excited state. Without the addition of any repumping light, we observe MOTs of metastable $^{46}$Ti, $^{48}$Ti, and $^{50}$Ti, the three stable nuclear-spin-zero bosonic isotopes of Ti. While MOTs can be observed when loaded directly from our Ti sublimation source, optical pumping of ground term atoms to the a$^5$F$_5$ state increases the loading rate by a factor of 120, and the steady-state MOT atom number by a factor of 30. At steady state, the MOT of $^{48}$Ti holds up to $8.30(26)\times10^5$ atoms at a maximum density of $1.3(4)\times10^{11}\mathrm{cm}^{-3}$ and at a temperature of $\qty{90(15)}{\micro\kelvin}$.  By measuring the decay of the MOT upon suddenly reducing the loading rate, we place upper bounds on the leakage branching ratio of the cooling transition ($\leq2.5\times 10^{-6}$) and the two-body loss coefficient ($\leq2\times10^{-10}\mathrm{cm}^3\mathrm{s}^{-1}$). Our approach to laser cooling Ti can be applied to other transition metals, enabling a significant expansion of the elements that can be laser cooled.
\end{abstract}

\maketitle

The magneto-optical trap (MOT)~\cite{raab_trapping_1987} serves as the starting point for a wide range of experiments on ultracold atoms and molecules. Such trapping uses two features of the particles being cooled.  The first is a near-cycling transition, in which an atom or molecule excited from a lower-energy state (the laser-cooling state) returns to that state with only a small leakage branching ratio to other states.  This feature allows one to capture atoms or molecules from high initial velocities through radiation pressure.  The second feature is a Zeeman shift of $\sigma^\pm$ optical transitions, which, in combination with magnetic field gradients, produces a restoring force that compresses the MOT to a small volume.  While MOTs have been realized for a minority of atomic elements, and also for a small selection of molecules, the ubiquitous application of MOTs has been limited by the absence of cycling optical transitions at optical wavelengths accessible to modern laser technologies.

We focus on the possibility of laser cooling and trapping transition-metal atoms.  The level structure for such atoms contains many metastable states, corresponding to different configurations of electrons among the valence $\mathrm{s}$ and $\mathrm{d}$ orbitals.  Radiative decay to these metastable states causes most optical transitions to be non-cycling, and thus unsuitable for laser cooling.  For example, resonant transitions out of the $\mathrm{3d^2 4s^2}$ $\mathrm{a}\tensor*[^3]{\mathrm{F}}{}$ ground term of atomic Ti allow cycling of only tens of photons before leakage occurs to other states.  However, it has been  pointed out that electronic configurations in which a single valence electron occupies the $n\mathrm{s}$ orbital may possess a cycling $n \mathrm{s}_{1/2} \rightarrow n\mathrm{p}_{3/2}$ transition~\cite{eustice_laser_2020}.  Laser cooling on such a transition should proceed as it does for D2 transitions in alkali atoms.  Laser cooling of iron follows this principle~\cite{crauwels_magneto-optical_2016}.

\begin{figure*}
    \centering
    \includegraphics[width=\linewidth]{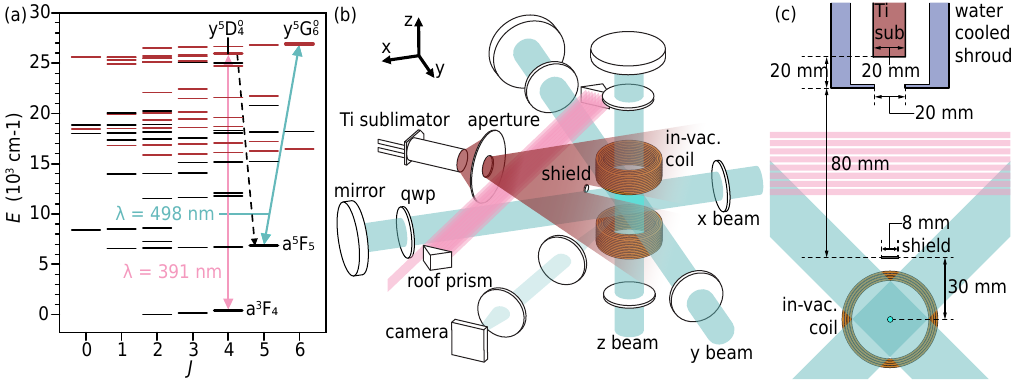}
    \caption{(a) Relevant energy levels of Ti~\cite{kramida_nist_1999, eustice_optical_2023} (black/red = even/odd parity). Transitions are identified with their wavelength $\lambda$: optical pumping light ($\lambda = 391$ nm, pink arrow) and laser-cooling light ($\lambda = 498$ nm, cyan arrow).
    (b) Diagram of the experimental apparatus. The Ti sublimator emits atomic Ti (maroon) through the aperture of a water-cooled shroud to form an atomic beam. Roof prisms pass optical pumping light through the atomic beam 8 times.  Three retroreflected Gaussian MOT beams ($1/e^2$ beam radius $w_0=\qty{15}{\mm}$, clipped to a diameter of \qty{22}{\mm}) are polarized in the $\sigma+$-$\sigma-$ configuration. In-vacuum coils (orange) produce the MOT spherical-quadrupole field, with a maximum axial gradient of \qty{19.4}{\gauss\per\cm}. A shield prevents hot atoms from directly colliding with the atoms trapped in the MOT. Fluorescence by the MOT of laser-cooling light, or transmission through the MOT of a resonant probe beam, is imaged onto a camera. An ion pump, omitted from the diagram for clarity, serves as a vacuum pump and a pressure gauge for the chamber.
    (c) A zoomed-in, top-down view of the experimental chamber. The components that generate the atomic beam (Ti sublimator, shroud, and shield), their key dimensions, and their distances  to the MOT position (cyan circle) are shown.
    }
    \label{fig:eng-and-exp-diagrams}
\end{figure*}

Here, we report the realization of laser slowing, cooling, and trapping following this scheme for atomic Ti.  In Ti, the laser-cooling $3\mathrm{d}^{3}(\tensor*[^4]{\mathrm{F}}{})4\mathrm{s}$ $\mathrm{a}\tensor*[^5]{\mathrm{F}}{_5}$ state is a long-lived metastable state.  Laser cooling is achieved by driving Ti from this state to the $3\mathrm{d}^{3}(\tensor*[^4]{\mathrm{F}}{})4\mathrm{p}$ $\mathrm{y}\tensor*[^5]{\mathrm{G}}{^o_6}$ excited state, with the transition wavelength being $\lambda = \qty{498}{\nano\meter}$ and the linewidth being $\Gamma = 2\pi \times \qty{10.8}{\mega\hertz}$~\cite{eustice_optical_2023} (Fig.~\ref{fig:eng-and-exp-diagrams}(a)).  This $J \rightarrow J+1$ transition permits the realization of a conventional ``type-I'' MOT~\cite{prentiss_atomic-density-dependent_1988}.  We observe MOTs for each of the three stable isotopes 
 --- \ce{^{46}Ti}, \ce{^{48}Ti}, and \ce{^{50}Ti} ---  that are free of hyperfine structure because of their $I=0$ nuclear spin.  Through time-of-flight measurements, we measure a temperature of $T = \qty{90 \pm 15}{\micro\kelvin}$ for laser-cooled atoms, below the Doppler limit of $\qty{270}{\micro\kelvin}$, indicating the presence of polarization-gradient cooling.  By measuring the loss of atoms from the MOT, we constrain the leakage branching ratio $\alpha$ on the laser-cooling transition to be $\leq2.5 \times 10^{-6}$, and the two-body loss coefficient $\beta$ to be $\leq\qty{2e-10}{\centi\meter^3\per\second}$.

We use a highly simplified apparatus for this first demonstration of Ti laser cooling.  As illustrated in Fig.~\ref{fig:eng-and-exp-diagrams}(b,c), we load a MOT directly with the emission of a commercially available Ti-sublimation vacuum pump that operates at approximately \qty{1350}{\kelvin}~\cite{schrott_atomic_2024}. At this temperature, only a small fraction (0.04\%) of the sublimated atoms are thermally excited to the laser-cooling state.  As demonstrated in Ref.~\cite{schrott_atomic_2024}, this fraction can be enhanced through optical pumping. Our system operates under a similar principle to a vapor-cell MOT~\cite{monroe_very_1990}, where only the atoms in the low-velocity tail of the Maxwell--Boltzmann distribution can be captured. While a standard vapor-cell MOT captures incident atoms from all directions onto the MOT capture area, our Ti-sub-loaded MOT captures atoms only from the small solid angle subtended by the Ti-sublimation source. Additionally, a small metal shield is placed in the atomic beam to block direct propagation from the Ti-sub pump to a few-mm sized region around the MOT center. This simple initial apparatus allows a MOT to be loaded only at a relatively slow loading rate (observed up to $\qty{5.7(3)e7}{\per\second}$). Nevertheless, the small leakage branching ratio and two-body loss coefficient revealed by this apparatus indicate that standard methods to increase the MOT loading rate, such as Zeeman slowing~\cite{phillips_laser_1982}, should allow for laser cooling and trapping large numbers of Ti atoms.

\begin{figure}
    \centering
    \includegraphics[width=\columnwidth]{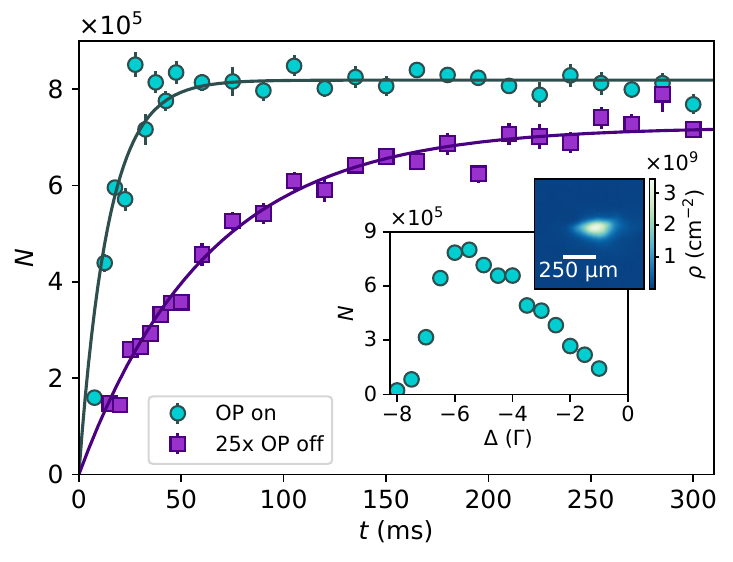}
    \caption{The number of atoms trapped in the MOT ($N$) after a loading time $t$ is shown. The cyan circles show the number of atoms loaded when the optical pumping (OP) light is turned on, and the purple squares show the number loaded when the optical pumping light is turned off, multiplied by a factor of 25 for visibility. A fit to the form $N(1-e^{-t/\tau})$ with $N$ and $\tau$ as free parameters allows us to determine the loading rate, $R=N/\tau$. Error bars reflect the statistical uncertainty in the data. The inset figure shows the variation of the number of atoms trapped in the MOT versus the laser detuning $\Delta$, with the largest MOTs appearing at $\Delta=-5.5\, \Gamma$. We show an example absorption image of such a MOT, with the pixel color corresponding to the column density of atoms.}
    \label{fig:loading_and_optimiz}
\end{figure}

In our setup, a small portion of the sublimated atoms travel forward directly onto the capture area of the MOT. This area is defined by the intersection of three mutually orthogonal, retro-reflected light beams with $1/e^2$-intensity beam radii of \qty{15}{\mm} that are further clipped to \qty{22}{\mm} diameters to reduce stray light. The beams intersect at the center of a spherical-quadrupole magnetic field, formed by in-vacuum coils, and located \qty{13}{\cm} from the sublimation source.  The intensity $I_z$ of the beam propagating along the quadrupole-field axis is $1.5\times$ smaller than the intensities $I_{x,y}$ of the other pair of beams.  Here, we followed the example of experiments on Er and Dy, other atoms for which the laser-cooling state has high $J$, where this intensity ratio produced larger and colder MOTs~\cite{berglund_sub-doppler_2007, youn_dysprosium_2010, youn_anisotropic_2010}.

Despite the low fraction of metastable atoms produced thermally at the sublimation temperature, we observe a MOT of Ti even when loading directly from the sublimated beam.  In Fig.~\ref{fig:loading_and_optimiz}, we show the number of atoms in the MOT, measured by imaging the atomic fluorescence onto a camera, as a function of time after switching on the MOT beams.  This fluorescence measurement is calibrated against resonant absorption imaging of the steady-state MOT numbers shown in the Fig.~\ref{fig:loading_and_optimiz} inset. For both measures, we assume a fully unpolarized atomic $m_J$ distribution~\cite{numberfootnote}.

We observe up to $2.8(2)\times10^4$ atoms loaded directly from the thermal beam at a loading rate of $\qty{4.6(2)e5}{\per\s}$.  For these data, the power in the three beams sent into the MOT is \qty{410}{\mW}, corresponding to $I_z = \qty{30}{\mW\per\cm^2}$; the axial spherical-quadrupole field gradient is $B^\prime = \qty{19.4}{\gauss\per\cm}$; and the light is detuned by $\Delta = -5.5 \, \Gamma = 2 \pi \times \qty{-59.4}{\MHz}$ from the laser-cooling transition resonance. Here, and unless stated otherwise, we report results for \ce{^{48}Ti}, which is the most naturally abundant isotope of Ti.  We emphasize the simplicity of this arrangement, requiring just a Ti-sub vacuum pump as an atom source and one single-frequency laser for cooling.

We enhance the loading rate and the atom number of the MOT by enriching the metastable-state population through optical pumping.   For this, a 391-nm-wavelength laser beam is passed 8 times, using roof prisms, through the forward emission of the Ti-sub pump.  This collimated \qty{15}{\mW} beam has a vertical (horizontal) beam-waist radius of \qty{12}{\milli\meter} (\qty{1.5}{\milli\meter}).  The light resonantly drives the $\mathrm{3d^2 4s^2}$ $\mathrm{a}\tensor*[^3]{\mathrm{F}}{_4} \rightarrow \mathrm{3d^2(\tensor*[^3]{\mathrm{P}}{})4s4p(\tensor*[^3]{\mathrm{P}}{^o})}$ $\mathrm{y}\tensor*[^5]{\mathrm{D}}{^o_4}$ transition, and pumps about 50\% of atoms from the $\mathrm{a}\tensor*[^3]{\mathrm{F}}{_4}$ state into the laser-cooling state~\cite{schrott_atomic_2024,opfootnote}.   Upon applying this optical pumping light, the MOT loading rate rapidly increases, reaching a maximum value of $\qty{5.7(3)e7}{\per\second}$ and a steady-state MOT number as high as $8.30(26)\times10^5$, at the same experimental settings as stated above.

We observe the MOT loading times, i.e.\ the $1/e$ relaxation time to the steady-state MOT atom number, of \qty{14(1)}{\ms} with and \qty{62(3)}{\ms} without optical pumping.  The difference in loading times indicates an enhanced loss of MOT atoms in the presence of optically pumped atoms. 
Based on our analysis of the lifetime of the MOT, discussed below, this increased loss rate cannot be explained by collisions between atoms trapped in the MOT. Rather, we posit that optical pumping generates many fast metastable-state atoms that cannot be captured in the small MOT region. These fast atoms do not have direct line-of-sight access to the MOT, owing to the aforementioned shield. However, unlike the ground-term atoms, fast-moving atoms that have been pumped to the metastable state could be redirected by optical and/or magnetic forces in the MOT volume and then collide with and expel atoms from the MOT.

Using absorption imaging, we examined the steady-state MOT number reached in our setup as a function of several MOT parameters: the MOT-light detuning $\Delta$ and power $P$, and the axial field gradient $B^\prime$.  Here, the Ti-sub temperature and settings for optical pumping were kept constant.  At each $\Delta$, we found that maximizing $P$ and $B^\prime$ within experimental bounds resulted in the largest number of trapped atoms. Having set those parameters, we found the MOT number increased with $|\Delta|$ up to $\Delta = -5.5\, \Gamma$ (see inset of Fig.~\ref{fig:loading_and_optimiz}).  The larger detuning presumably increases the capture velocity of the MOT, increasing the loading rate and steady-state atom number.

We determine the size of the MOT by fitting absorption images with a two-dimensional Gaussian distribution. We find the $1/e^2$ radii are $\sigma_z=\qty{178(14)}{\micro\meter}$ and $\sigma_r=\qty{90(6)}{\micro\meter}$ for the axial and imaged radial dimensions. Assuming cylindrical symmetry about the quadrupole-field axis, we determine the maximum density of atoms to be \qty{1.3(4)e11}{\cm^{-3}}. This density is comparable to the highest densities observed in bright alkali MOTs~\cite{townsend_phase-space_1995}.

As mentioned above, the $J \rightarrow J+1$ nature of the laser-cooling transition should support sub-Doppler polarization-gradient cooling for red-detuned light.  To test this expectation, we performed time-of-flight measurements of the temperature of atoms in the MOT.  For this, we suddenly switch off both the MOT light and the magnetic field gradients, and allow a variable time of flight before measuring the atomic distribution through fluorescence imaging.  From these measurements, we determine the MOT temperature to be \qty{90(15)}{\micro\kelvin} for the MOT parameters described above.

Let us compare this temperature to the laser-cooling Doppler limit, which is conventionally defined by the one-dimensional result: $T_D(I_{\mathrm{tot}},\Delta) = (\hbar \Gamma/4 k_{\mathrm{B}}) \left[(1 + I_{\mathrm{tot}}/I_0 + (2 \Delta/\Gamma)^2)/(- 2\Delta/\Gamma)\right]$~\cite{lett_optical_1989}.  Here, $I_{\mathrm{tot}}$ is the combined optical intensity of the cooling beams and $I_0=\qty{11.4}{\mW}$ is the saturation intensity of the stretched-state cooling transition.
At the settings of our MOT ($\Delta = -\Gamma/2$ and $I_{\mathrm{tot}}\ll I_0$), the predicted Doppler temperature is $T_D(21I_0, -5.5\, \Gamma) = \qty{1.6}{\milli\kelvin}$, considerably higher than the conventional Doppler limit, $T_D(I\ll I_0, -\Gamma/2)=\qty{270}{\micro\kelvin}$. 
The observed MOT temperature is lower than either of these limits, demonstrating the effectiveness of sub-Doppler polarization-gradient cooling throughout the MOT.  The observation that polarization-gradient cooling is not impaired by the presence of magnetic field gradients in a Ti MOT is similar to what is observed for Er and Dy~\cite{berglund_sub-doppler_2007,youn_anisotropic_2010}.  The effectiveness in all cases may be ascribed to the Land\'e $g_J$ factors being nearly equal in the lower- and the higher-energy states on the laser-cooling transition, with ratios of 0.945, 0.997, and  0.983 in Ti, Er, and Dy, respectively.  It is likely that even lower Ti MOT temperatures can be achieved with better control over the intensity balance between MOT beams, and also by transiently reducing the intensity or increasing the detuning of MOT beams as implemented in other laser cooling experiments.

\begin{table}[]
    \centering
    \begin{tabular}{lcccc}
    \hline
        Isotope & Natural & $N$ & $\delta f_{\mathrm{cool},A}$ & $\delta f_{\mathrm{OP},A}$ \\ 
        & Abund.~\cite{meija_isotopic_2016} & & [MHz] & [MHz] \\\hline \hline
        \ce{^{46}Ti}  & $0.08249(21)$ & $1.80(8)\!\times\!10^5$ & $-703(2)$ &  $-900(2)$  \\
        \ce{^{48}Ti} & $0.73720(22)$ & $8.30(26)\!\times\!10^5$ & --- & --- \\
        \ce{^{50}Ti} & $0.05185(13)$ & $1.22(7)\!\times\!10^5$ & $673(2)$ & $880(2)$\\
    \end{tabular}
    \caption{
    Overview of the MOTs of the three trapped Ti isotopes. The natural abundance, the number of atoms trapped in the MOT ($N$), the isotope shifts of the laser-cooling transition ($\delta f_{\mathrm{cool},A} = f_{\mathrm{cool},A}-f_{\mathrm{cool},48}$) and the optical pumping transition ($\delta f_{\mathrm{OP},A}$) are given. The isotope shifts of the cooling transition were measured by absorption spectroscopy of the MOT, while the shifts of the optical pumping transition were measured by saturated absorption spectroscopy in a hollow cathode lamp as in Ref.~\cite{neely_isotope_2021}. We note that there is a \qty{13}{\MHz} discrepancy between our previous measurement of $\delta f_{\mathrm{cool,}46}$ and the value measured in the MOT~\cite{neely_isotope_2021}.
    }
    \label{tab:isotope_properties}
\end{table}

By tuning the optical pumping and MOT light to account for isotope shifts, we also observe MOTs for the other two stable Ti isotopes with $I=0$, \ce{^{46}Ti} and \ce{^{50}Ti}.  We note that the relative sizes of MOTs for these three isotopes do not precisely follow their natural abundance (see Tab.~\ref{tab:isotope_properties}).  We believe this difference is explained by the increased MOT loss rate in the presence of an optically pumped beam, as described above. We are unable to observe MOTs of the two fermionic isotopes, \ce{^{47}Ti} and \ce{^{49}Ti}. Hyperfine structure on the optical-pumping and laser-cooling transitions in these isotopes will require additional repump light to maintain the population in the stretched (maximum total spin) laser-cooling state.

Laser cooling of a complex atom could face limitations that are not present when laser cooling simpler atoms.  For one, in Ti, the laser-cooling transition is not strictly closed.  Below the $\mathrm{y}\tensor*[^5]{\mathrm{G}}{^o_6}$ excited state, there lie several even-parity states other than the laser-cooling state to which spontaneous emission is dipole-allowed: the $\mathrm{a}\tensor*[^3]{\mathrm{G}}{_5}$, $\mathrm{a}\tensor*[^3]{\mathrm{H}}{_{5,6}}$, and $\mathrm{a}\tensor*[^1]{\mathrm{H}}{_5}$ states.  An atom scattered into one of these other long-lived metastable states no longer scatters laser-cooling light, and thus can be considered lost from the MOT. The transitions to these states are spin-forbidden, and as a result, the total emission branching ratio to them is predicted to be low ($\alpha_{\mathrm{theory}} = 1.1(5) \times 10^{-6}$~\cite{eustice_optical_2023}).  This leakage branching ratio has not yet been measured. 

Another possible mechanism for the loss of atoms from the MOT is two-body collisions.  Light-induced collisions may occur between an $\mathrm{a}\tensor*[^5]{\mathrm{F}}{_5}$ and a $\mathrm{y}\tensor*[^5]{\mathrm{G}}{^o_6}$ atom.  Additionally, an exothermic collision can occur between two $\mathrm{a}\tensor*[^5]{\mathrm{F}}{_5}$ atoms, leading to fine-structure relaxation or decay to the electronic ground term.  In either case, the collision can release sufficient energy to expel both atoms from the MOT.

\begin{figure}
    \centering
    \includegraphics[width=\linewidth]{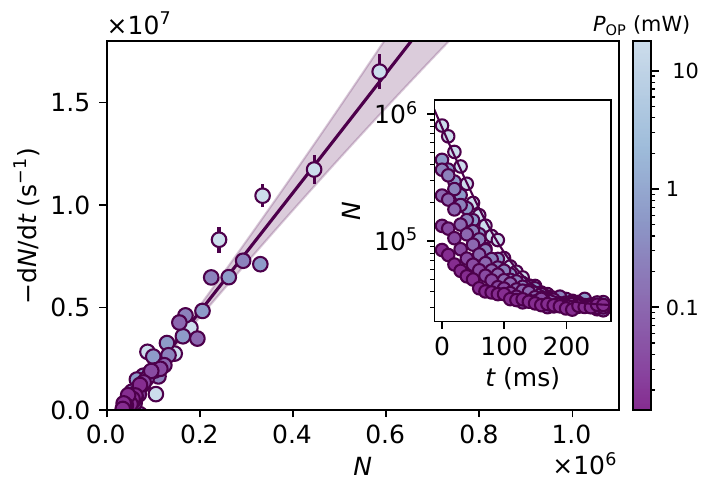}
    \caption{MOT atom loss rate $\mathrm{d}N/\mathrm{d}t$ vs. MOT atom number $N$.  The line shows a fit to Eq.~\ref{eq:mot_diff_eq}; shading shows error bounds.  The predominantly linear relationship between $\mathrm{d}N/\mathrm{d}t$ and $N$ indicates that two-body loss mechanisms are not significant in our MOT.  Inset: Time traces of $N(t)$. Varying the optical pumping power $P_{\mathrm{OP}}$ (indicated by marker color) adjusts the initial atom number by about 10-fold without changing the trap volume. $\mathrm{d}N/\mathrm{d}t$ in the figure is obtained from finite differences in the time traces.  Data agree well with a fit to pure one-body decay plus a residual loading rate (one fit shown as a solid line in the inset).
    }
    \label{fig:lifetime-measurement}
\end{figure}

\begin{figure}
    \centering
    \includegraphics[width=\linewidth]{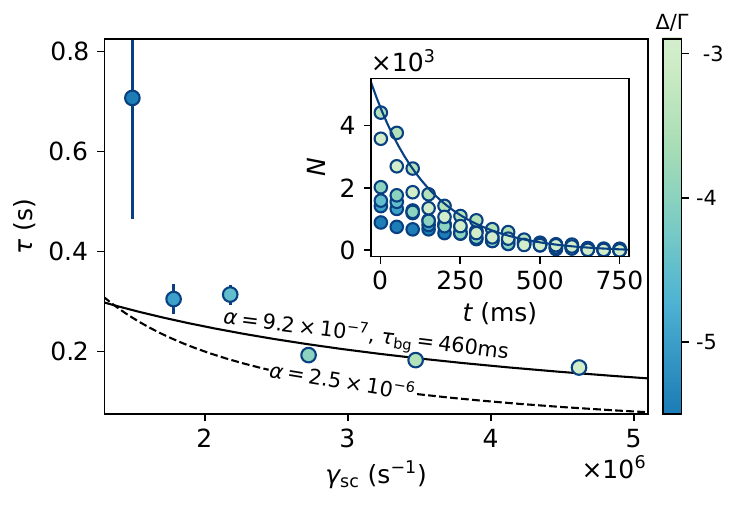}
    \caption{The one-body lifetime $\tau$ is measured as a function of the scattering rate $\gamma_{\mathrm{sc}}$ in the MOT at low sublimator temperatures. Error bars reflect fit errors. We vary $\gamma_{\mathrm{sc}}$ by changing $\Delta$ (indicated by marker color) while keeping the MOT beam power ($P$) and field gradient ($B'$) fixed.
    The solid line shows a fit of the data to $\tau^{-1}=\alpha \gamma_{\mathrm{sc}}+\tau_{\mathrm{bg}}^{-1}$, indicating $\alpha = 9.2(3.5)\times10^{-7}$ and $\tau_\mathrm{bg} = \qty{460(250)}{\ms}$.  
    The dashed line shows a curve of $\tau^{-1}=\alpha \gamma_{\mathrm{sc}}$ with $\alpha=2.5\times10^{-6}$, which serves as an upper bound on $\alpha$ after accounting for systematic uncertainties. 
    Insert: time traces from which the lifetimes were extracted; exponential-decay fit for one trace.}
    \label{fig:one-body-lifeitme}
\end{figure}

We constrain both these one- and two-body loss processes by filling the MOT using the optically pumped source, abruptly extinguishing the optical pumping light, and tracking the subsequent decay of the MOT atom number by fluorescence imaging.  We perform these measurements at two experimental settings.  In the first (Fig.~\ref{fig:lifetime-measurement}), we load the MOT to large atom numbers by operating the Ti sublimation source at a high temperature.  The decay of these large MOTs serves to constrain the two-body loss rate.  In the second setting (Fig.~\ref{fig:one-body-lifeitme}), we load the MOT while operating the Ti sublimation source at a low temperature, where the vacuum pressure in the MOT chamber is reduced due to lower outgassing from the colder vacuum chamber walls.  Measurements at this setting serve to constrain the leakage branching ratio of the laser-cooling transition.

We model the decay of the MOT atom number $N$ in either setting using a rate equation of the form~\cite{prentiss_atomic-density-dependent_1988}
\begin{equation}
    \label{eq:mot_diff_eq}
    \frac{\mathrm{d}N}{\mathrm{d}t}=R_{\mathrm{th}}-\frac{N}{\tau} -\tilde{\beta}N^2.
\end{equation}
Here, $R_{\mathrm{th}}$ is the loading rate of atoms that are thermally excited to the $\mathrm{a}\tensor*[^5]{\mathrm{F}}{_5}$ state without optical pumping and $\tau$ is the lifetime under one-body loss. Our rate equation includes a scaled two-body loss coefficient $\tilde{\beta} = \beta \langle n^2 \rangle/N^2$, where $\beta n^2$ is the two-body loss rate at a uniform density $n$, and where brackets denote a spatial average over the MOT volume. Assuming the MOT has a Gaussian spatial distribution with cylindrical symmetry and constant $1/e^2$ density radii in the axial and radial dimensions of $\sigma_{z,r}$, $\tilde{\beta}$ has the form $\tilde{\beta} = \beta / (8\pi^{3/2} \sigma_r^2\sigma_z)$.

In Fig.~\ref{fig:lifetime-measurement}, we show the MOT atom loss rate $\mathrm{d}N/\mathrm{d}t$, as a function of $N$, determined from several time traces of the decay of the MOT.  Here, the MOT atom number was adjusted by filling the MOT to a steady state with a different powers of optical pumping light before that light was extinguished at $t=0$.  Following Eq.~\ref{eq:mot_diff_eq}, two-body loss would appear as a quadratic dependence of $\mathrm{d}N/\mathrm{d}t$ with $N$ and a deviation from exponential decay of the atom number.  From fitting our data, we find $\tilde{\beta} = \qty[parse-numbers=false]{5(8)\times10^{-6}}{\per\second}$, consistent with zero. Using our measurements of $\sigma_z=\qty{178(14)}{\micro\meter}$ and $\sigma_r=\qty{90(6)}{\micro\meter}$, we establish an upper bound on the two-body loss rate of $\beta\leq\qty{2e-10}{\cm^3\per\second}$.

Fixing $\beta=0$, we measure the one-body lifetime to be $\tau=\qty{39(3)}{\ms}$. We attribute this short lifetime mainly to collisions with background gas atoms.
With the Ti-sub source running at the high temperature setting used here, we estimate the background gas pressure at the MOT to be \qty{4e-8}{\mbar} based on readings from the chamber's ion pump and knowledge of the chamber geometry.

To place constraints on the leakage branching ratio $\alpha$, we reduce the sublimator temperature. This reduces the flux of Ti atoms by a factor of $\approx1000$ and the background gas pressure to about \qty{3e-9}{\mbar}. Under these conditions, we observed a significantly increased one-body lifetime, as shown in the inset of Fig.~\ref{fig:one-body-lifeitme}. As the number of atoms trapped in the MOT and thus the MOT density are greatly reduced, we are able to neglect the effects of two-body loss entirely. We consider that leakage from the laser-cooling transition should cause atoms to be lost from the MOT at rate $\tau_{\alpha}^{-1} = \alpha \gamma_\mathrm{sc}$, with $\gamma_\mathrm{sc}$ being the rate at which an atom scatters laser-cooling light. We model the other sources of one-body loss as a constant background loss rate $\tau_{\mathrm{bg}}^{-1}$, leading to a total loss rate $\tau^{-1} = \tau_\alpha^{-1} + \tau_{\mathrm{bg}}^{-1}$.  To measure the leakage from the laser-cooling transition on the MOT lifetime, we vary $\gamma_{\mathrm{sc}}$ by tuning $\Delta$ while keeping $P$ and $B^\prime$ fixed. The results are plotted in Fig.~\ref{fig:one-body-lifeitme}. We note that there is a large, non-linear systematic uncertainty in our determination of $\gamma_{\mathrm{sc}}$, dominated by uncertainty in the linewidth of the laser-cooling transition and in the intensity of the MOT beams. For clarity, this uncertainty is not shown in Fig.~\ref{fig:one-body-lifeitme}. Fitting the model for $\tau$, we find $\alpha= 9.2(3.5)_{\mathrm{stat}} \left(^{+7.7}_{-4.8}\right)_{\mathrm{sys}} \times10^{-7}$, in agreement with theoretical predictions~\cite{eustice_optical_2023}. Noting that the MOT detuning, $\Delta$, may influence the background loss rate, $\tau_\mathrm{bg}$, through a number of mechanisms --- for example by changing the trap depth of the MOT --- we place a conservative upper bound on $\alpha$ by modeling the loss from the MOT as being solely due to leakage (i.e., $\tau=\tau_\alpha$). Accounting for systematic uncertainties in $\gamma_{\mathrm{sc}}$, we constrain $\alpha\leq2.5\times10^{-6}$.

The low atomic mass of Ti and the large Land\'{e} g-factor ($g_J = 1.41$) in the laser-cooling state imply that metastable Ti atoms can be trapped purely magnetically (i.e., without laser-cooling light) in the modest spherical-quadrupole field imposed in the MOT. To demonstrate this, we suddenly extinguish the laser cooling and optical pumping light. For atoms with $m_J>0$, the spherical quadrupole field of the MOT produces a trapping potential with peak trap acceleration (trap depth) of $\qty{31.8}{\meter\per\second^2}$ (\qty{1.3}{\milli\kelvin}) for atoms with $m_J=1$ and $\qty{159}{\meter\per\second^2}$ (\qty{8.9}{\milli\kelvin}) for atoms with $m_J=5$. As the trapping forces are stronger than gravity and the trap depths are deeper than the MOT temperature, we expect to load all atoms with $m_J>0$ into the magnetic trap. After a time delay, we measure the number of atoms still localized within the trapping region by fluorescence imaging. We observe up to $5.1(5)\times10^5$ magnetically trapped atoms, representing 60\% of the MOT atom number, at a peak density of \qty{2.5(4)e9}{\cm^{-3}}.  The magnetic trap lifetime, which ranged between \qty{0.34(11)}{\second} and \qty{0.11(4)}{\second} as the chamber pressure was increased from \qty{2e-8}{\mbar} to \qty{5e-8}{\mbar}, appears to be vacuum limited.

In conclusion, we realize the first laser cooling and trapping of the three bosonic isotopes of Ti. Using an optically pumped thermal atomic beam as the source of Ti atoms, we trap $8.30(26)\times10^{5}$ atoms of the most abundant isotope, \ce{^48Ti}, at \qty{90(15)}{\micro\kelvin}.  We are also able to load a fraction of these atoms into a spherical-quadrupole magnetic trap. While the MOT lifetime is limited by high vacuum pressure and collisions with fast Ti atoms, we are able to place constraints on the leakage branching ratio of the laser-cooling transition ($\alpha \leq 2.5\times10^{-6}$) and on the two-body loss coefficient in the MOT ($\beta \leq \qty{2e-10}{\cm^3\per\second}$). Separating the MOT further from the Ti-sub source, e.g., by using a Zeeman slower, should lead to increased loading rates and lifetimes. Ti also possesses a second laser-cooling transition ($\lambda=\qty{1040}{\nm}, \Gamma=2\pi\times\qty{11}{\kHz}$, $T_{\mathrm{D}}=\qty{250}{\nano\kelvin}$)~\cite{eustice_laser_2020,kramida_nist_1999} which may allow the implementation of all-optical cooling to below \qty{1}{\micro\kelvin}. The method of laser cooling demonstrated here for Ti may be applicable to a number of other transition metal atoms~\cite{eustice_laser_2020}, opening the door to exploring a wider range of quantum gases.

\begin{acknowledgments}
We thank Marianna Safronova, Dmytro Filin, Sergey Porsev, and Charles Cheung for their calculation of transition rates and branching ratios. We thank Isaac Pope and Andrew Neely for their work on isotope shift spectroscopy. This work is supported by a collaboration between the US DOE and other Agencies.  This material is based upon work supported by the U.S. Department of Energy, Office of Science, National Quantum Information Science Research Centers, Quantum Systems Accelerator.  Additional support is acknowledged from the ONR (Grant Nos.~N00014-20-1-2513 and N00014-22-1-2280), the ARO (Grant No.~W911NF2010266), the NSF (PHY-2012068 and the QLCI program through Grant No.~OMA-2016245), and the Heising-Simons Foundation.
\end{acknowledgments}

\bibliography{references,%
            footnotes,%
            manual_refs
            }

\begin{thebibliography}{18}
\bibitem [{num()}]{numberfootnote}%
  \BibitemOpen
  \href@noop {} {}\bibinfo {note} {We ascribe a 15\% systematic uncertainty in all reported atom numbers owing to a combination of uncertainty about the $m_J$ distribution of the atoms (assumed uniform) and an uncertain imaging magnification. There is a additional 5\% uncertainty in atom numbers measured by fluorescence due to calibration uncertainties. We only report statistical uncertainties for the remainder of the report, unless otherwise noted.}\BibitemShut {Stop}%
\end{thebibliography}

\end{document}